\documentclass[aps,prl,twocolumn,superscriptaddress,amsmath,showpacs]{revtex4}

\usepackage{graphicx}

\begin{document}


\title{Andreev bound states and tunneling characteristics of a \\
non-centrosymmetric superconductor}



\author{C. Iniotakis}
\affiliation{Institute for Theoretical Physics, ETH Zurich, 8093 Zurich, Switzerland}
\author{N. Hayashi}
\affiliation{Institute for Theoretical Physics, ETH Zurich, 8093 Zurich, Switzerland}
\author{Y. Sawa}
\affiliation{Department of Applied Physics, Nagoya University, Nagoya, 464-8603, Japan}
\author{T. Yokoyama}
\affiliation{Department of Applied Physics, Nagoya University, Nagoya, 464-8603, Japan}
\author{U. May}
\affiliation{Institute for Theoretical Physics, ETH Zurich, 8093 Zurich, Switzerland}
\author{Y. Tanaka}
\affiliation{Department of Applied Physics, Nagoya University, Nagoya, 464-8603, Japan}
\author{M. Sigrist}
\affiliation{Institute for Theoretical Physics, ETH Zurich, 8093 Zurich, Switzerland}
\affiliation{Department of Physics, Kyoto University, Kyoto 606-8502, Japan}


\date{\today}

\begin{abstract}
The tunneling characteristics of planar junctions between a normal metal and a non-centro\-symmetric superconductor like CePt$_3$Si are examined. It is shown that the superconducting phase with mixed parity can give rise to characteristic zero-bias anomalies in certain junction directions.
The origin of these zero-bias anomalies are Andreev bound states at the interface. The tunneling characteristics 
for different directions allow to test the structure of the parity-mixed pairing state.
 \end{abstract}

\pacs{74.20.Rp, 74.70.Tx, 74.45.+c}


\maketitle


Since the early sixties tunneling spectroscopy has played an important 
role in gathering information
about the gap function of conventional superconductors \cite{Giaever}. 
In the context of unconventional superconductivity 
tunneling appeared as a tool to probe the internal phase structure of the 
Cooper pair wave functions. Surface states with sub-gap energy, known as Andreev bound states,
provide channels for resonant tunneling leading to so-called zero-bias anomalies.
Zero-bias anomalies observed in high-temperature superconductors showed the presence
of zero-energy bound states at the surface, giving strong evidence for $d$-wave pairing \cite{Hu, 
Tanaka95, Buchholtz, Bruder, KashiwayaReport}.
Similarly the tunneling spectrum observed in Sr$_2$RuO$_4$ is consistent with
the existence of chiral surface states as expected for a chiral $p$-wave superconductor
 \cite{Yamashiro,Honerkamp,Matsumoto,Laube,Mao}.  Thus quasiparticle tunneling has emerged as an
important phase sensitive probe for unconventional superconductors.

The recent discovery of superconductivity in the heavy Fermion compound 
CePt$_3$Si has motivated intense experimental and theoretical studies, since
its  crystal symmetry is lacking a center of inversion \cite{Bauer, Frigeri, Samokhin, BauerReport, Hayashi}.
 It was shown that the
presence of antisymmetric spin-orbit coupling in such a material could be
responsible for a number of intriguing phenomena, such as the rather high
upper critical field exceeding the standard paramagnetic limit \cite{Bauer,Frigeri}.
The absence of inversion symmetry yields a classification scheme of the pairing
symmetry, different from the standard distinction between even- and odd-parity states.
Actually, the pairing states in these non-centrosymmetric superconductors can be
viewed as states of mixed parity which do  not have a definite spin-singlet
or spin-triplet configuration, either. For CePt$_3$Si various experiments can be interpreted in
a way that the pairing state has the highest possible symmetry, but would develop  line nodes in the gap
due to a mixing of  the $s$-wave and a $p$-wave component.
This state belongs to the $A_1$-representation of the generating tetragonal point group 
$C_{4v}$ for CePt$_3$Si. 
While thermodynamic quantities give good evidence for this behavior, it would 
be desirable to have additional proof for this kind of state through a phase sensitive
test. The characteristics of quasiparticle tunneling may provide such a means. Despite the
fact that the parity mixing $A_1$-phase has the full symmetry of the system, it gives rise to  characteristic non-trivial
features in the tunneling spectra (cf. also  \cite{Yokoyama}). 

In the present study, we address the problem of possible Andreev bound
states  at the boundaries of a non-centrosymmetric 
superconductor like CePt$_3$Si and demonstrate that quasiparticle tunneling could
give important information on the gap structure of such a material.
The theoretical framework for our calculations is given by the quasiclassical Eilenberger 
theory \cite{Eilenberger,Larkin,Serene}. Our starting point is 
the  quasiclassical bulk propagator $\check{g}_B$ in a non-centrosymmetric 
superconductor with antisymmetric Rashba-type spin-orbit coupling, 
which has already been derived in Ref. \cite{Hayashi}:
\begin{equation}
\label{PropBulk}
{\check g}_B =-i\pi \left( \begin{array}{cc} 
\hat g & i\hat f \\  -i {\hat {\bar f}}  & -{\hat {\bar g}} \end{array} \right) ,
\end{equation}
where the upper two components are given as
\begin{eqnarray}
\label{gBulk}
\hat g &=&\frac{\omega_n}{\sqrt{\omega_n^2+|\Delta_{+} |^2}} \hat{\sigma}_{+}  +\frac{\omega_n}{\sqrt{\omega_n^2+|\Delta_{-} |^2}} \hat{\sigma}_{-} \\
\hat f &=&\left( \frac{\Delta_+}{\sqrt{\omega_n^2+|\Delta_+ |^2}} \hat \sigma_+ +\frac{\Delta_{-}}{\sqrt{\omega_n^2+|\Delta_{-} |^2}} \hat \sigma_{-} \right) i\hat \sigma_y
\label{fBulk}
\end{eqnarray}
and similar relations hold for the lower components ${\hat {\bar f}} $ and ${\hat {\bar g}}$.
On the right hand side of Eqs. (\ref{gBulk}) and (\ref{fBulk}), $\omega_n=(2n+1)  \pi k_B T$
 denotes the Matsubara frequency, and both the gap amplitudes $\Delta_\pm$ 
and the spin matrices  $\hat{\sigma}_\pm$ are $\hat {\mathbf k}$-dependent. 
For a spherical 
Fermi surface parametrized in the standard way by 
$\hat {\mathbf k}=(\cos\phi\sin\theta,\sin\phi\sin\theta,\cos\theta)$, the
gap amplitudes are
\begin{equation}
\label{Delta12}
\Delta_\pm=\psi \pm \Delta \sin\theta,
\end{equation}
while the spin matrices may be written as
\begin{equation}
\label{Sigma12}
\hat{\sigma}_\pm=\frac{1}{2}
\left( 
\begin{array} {cc} 1 & \mp i e^{-i \phi} \\ \pm i e^{i \phi}  & 1 \end{array}
\right).
\end{equation}
The variables $\psi$ and $\Delta$ are taken to be real
and denote the gap amplitudes of the $s$- and $p$-wave
components of the parity-mixed superconducting state.
Altogether, this result is interpreted as a superconducting phase 
on two split spherical Fermi surfaces, 
each of them having a nontrivial gap function 
with amplitudes given by 
Eq. (\ref{Delta12}) and complex spin structures
according to Eq. (\ref{Sigma12}). For technical simplicity we assume here that the
splitting is small, such that both Fermi surfaces have identical size and
a description by a one-band formalism is possible as 
described by Eqs. (\ref{PropBulk})-(\ref{Sigma12}). A generalization to a really split 
Fermi surface is straightforward, but requires a more extended notation
without giving additional insights.

In order to finally include the boundary effects, it is convenient
to use the Riccati-parametrization of the quasiclassical theory \cite{Schopohl,Eschrig}.
Within this technique, the full quasiclassical 
propagator $\check g$ in combined particle-hole and spin space may
be written in terms of two coherence functions 
$\gamma$ and $\tilde \gamma$ in the following way \cite{Eschrig}: 
\begin{equation}
\label{PropEschrig}
\check g=-i\pi \left( \begin{array}{cc} (1-\gamma \tilde\gamma)^{-1}&0\\0&(1-\tilde \gamma \gamma)^{-1} \end{array}  \right) 
\left( \begin{array}{cc} 1+\gamma \tilde\gamma & 2\gamma \\ -2\tilde \gamma & -1-\tilde \gamma \gamma \end{array}  \right) 
\end{equation}
The coherence functions $\gamma, \tilde \gamma$ themselves   
are $2 \times 2$-spin matrices, which contain local information 
about the particle-hole coherence for a given 
$\hat{ \mathbf{k}}$-direction. Particularly, $\gamma_B$ and $\tilde{\gamma}_B$ in the bulk of 
a non-centrosymmetric superconductor  
can be obtained  just by comparing the known result
of Eqs. (\ref{PropBulk})-(\ref{fBulk}) with the general form in Eq. (\ref{PropEschrig}), leading to:
\begin{subequations}
\begin{eqnarray}
\gamma_B &=& i (1+\hat{g})^{-1} \hat{f} \\
{\tilde \gamma}_B &=& i \hat f^{-1} (1-\hat g),
\end{eqnarray}
\end{subequations}
which  can be written as
\begin{subequations}
\begin{eqnarray}
\gamma_B &=&-(\gamma_+ {\hat \sigma}_+ +\gamma_{-}{\hat \sigma}_{-})\hat \sigma_y \\
{\tilde \gamma}_B &=& \hat \sigma_y (\gamma_+ {\hat \sigma}_+ +\gamma_{-}{\hat \sigma}_{-}), 
\end{eqnarray}
\end{subequations}
where we define
\begin{equation}
\gamma_{\pm} =\frac{\Delta_{\pm}}{\omega_n +\sqrt{\omega_n^2+|\Delta_{\pm} |^2}}  .
\end{equation} 
The advantage of this Riccati formalism is, that spatial inhomogeneities
such as the  boundaries   
can be taken into account in a straightforward way \cite{Zaitsev,Shelankov,Eschrig}. 

In the vicinity of the boundary the pair potential can develop a spatial dependence and
deviate from the bulk form. 
Since we focus mainly on the qualitative aspects, however, we ignore this effect
in the following and assume the gap is constant throughout the superconductor. 
The usual procedure within the Riccati formalism,
namely solving differential equations of the Riccati type
numerically for given $\hat{ \mathbf{k}}$-direction, is unnecessary in this case. 
The presence of the surface enters through rather simple boundary conditions 
for the coherence functions \cite{Eschrig}. For an impenetrable interface, the coherence
functions $\gamma_S,\tilde{\gamma}_S$ at the surface of the superconductor
are given analytically in terms of the bulk coherence functions:
\begin{eqnarray}
\label{CohFunc}
\gamma_S({\hat{ \mathbf{k}}}_{\textrm{out}})= \gamma_B ({\hat{ \mathbf{k}}}_{\textrm{in}}) & \quad  &
\tilde \gamma_S({\hat{ \mathbf{k}}}_{\textrm{out}})={\tilde \gamma}_B ({\hat{ \mathbf{k}}}_{\textrm{out}}).
\end{eqnarray} 
Here ${\hat{ \mathbf{k}}}_{\textrm{in}}$ and ${\hat{ \mathbf{k}}}_{\textrm{out}}$ denote 
the incoming and outgoing directions of the wave vector, which are connected through the
condition of specular scattering.
Now the full quasiclassical propagator 
${\check g}_S$ at the boundary is easily 
generated from the coherence functions
according to Eq. (\ref{CohFunc}). From  ${\check g}_{S,11}$ we obtain, for example, 
the angular resolved local density of states at the boundary of
the non-centrosymmetric superconductor:
\begin{eqnarray}
\label{NAngular} 
&&N(E,\hat{\mathbf{k}})\\&=&N_0 \frac{1}{2} \textrm{Re}\left[ \textrm{tr} \left. \left( (1-\gamma_S \tilde\gamma_S)^{-1} (1+\gamma_S \tilde\gamma_S) \right) \right|_{i \omega_n \rightarrow E+i 0^+}  \right],
\nonumber
\end{eqnarray}
where $E$ denotes the quasiparticle energy and $N_0$ is the density of states 
of the normal state. With $\langle ... \rangle$ denoting an average 
over the Fermi surface,
the local density of states is  just given as
$N(E)=\langle N(E,\hat{\mathbf{k}})\rangle$. Note, that at the 
boundary it is sufficient to average over half of the Fermi sphere
only, i.e. we can restrict ourselves to quasiparticles traveling in ${\hat{ \mathbf{k}}}_{\textrm{out}}$ 
directions. 

Eventually, we also obtain the normalized differential tunneling conductance 
for a normal metal/non-centrosymmetric superconductor junction
at low temperatures $T \ll T_c$. 
To first order of the interface transmission probability, it can be generated
from  the angular resolved local density of states of 
Eq. (\ref{NAngular}) 
in the following way
\cite{Barash}:
\begin{equation}
\label{GFormel}
G(eV)=\langle \cos\alpha \ D(\alpha) N(eV,\hat{\mathbf{k}})\rangle \langle \cos\alpha \ D(\alpha)  \rangle^{-1}.
\end{equation}
The function $D(\alpha)$ denotes
the transmission coefficient of the interface depending on 
the angle of incidence $\alpha$, which is related to the normal vector $\mathbf{n}$ 
of the boundary by 
$\cos\alpha= \mathbf{n}\cdot{\hat{ \mathbf{k}}}_{\textrm{out}}$.
In accordance with the result for an interface of the $\delta$-function 
type \cite{Bruder}, the transmission coefficient is 
taken to be
\begin{equation}
D(\alpha)=\frac{D_0 \cos^2 \alpha}{1-D_0 \sin^2 \alpha},
\end{equation}
where the parameter $D_0$  specifies the 
maximum transmission coefficient for the angle of normal incidence  $\alpha=0$.

\begin{figure}[t]
\includegraphics[width=0.85 \columnwidth]{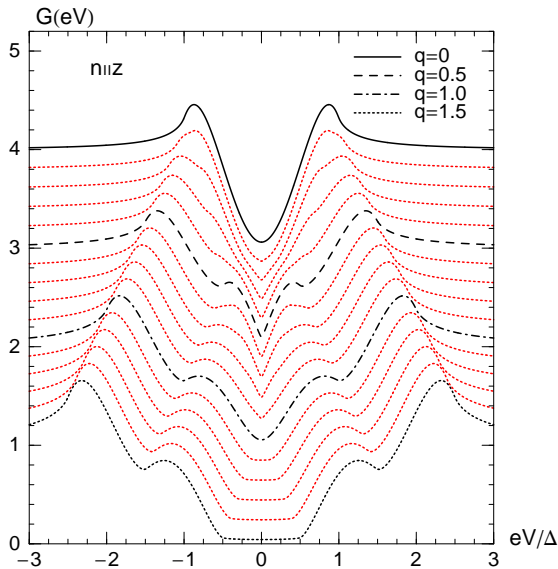}
\caption{\label{Fig01}
Normalized low temperature conductance $G(eV)$ for a normal 
metal/non-centrosymmetric superconductor junction
with boundary orientation $\mathbf{n}\parallel \mathbf{z}$. 
The ratio $q$ of the gap amplitudes is varied from $0$ to $1.5$ in 
steps of $0.1$. Between two subsequent curves, a vertical  offset
of $0.2$ has been used for clarity of the plot.
The transparency of the interface is set to $D_0=0.1$. 
}
\end{figure}

\begin{figure}[t]
\includegraphics[width=0.9 \columnwidth]{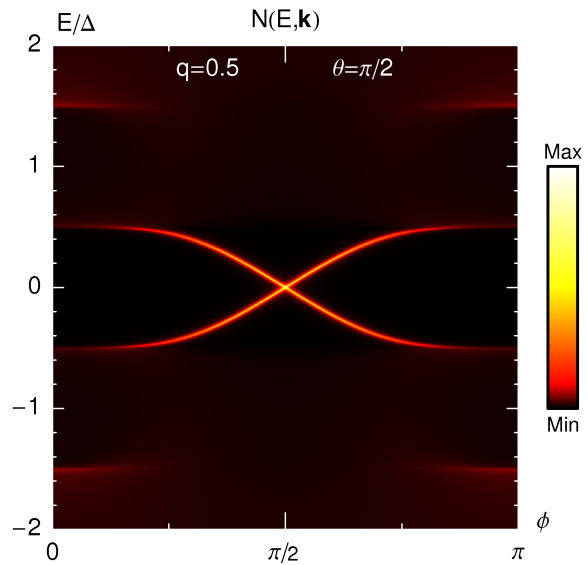}
\caption{\label{Fig02}
(\textit{color online}) 
Typical angular resolved density of states $N(E,\hat{\mathbf{k}})$ at the boundary of
a non-centrosymmetric superconductor with normal vector $\mathbf{n} \parallel \mathbf{y}$. 
For this example, we set $q=0.5$ and $\theta=\pi/2$, i.e. we deal
with $\hat{\mathbf{k}}$-vectors of the $x,y$-plane only. 
Bright colors represent high spectral weight. 
The branches of low-energy Andreev bound states  according to
Eq. (\ref{Branch}) are clearly visible.    
}
\end{figure}

For the surface with normal vector $\mathbf{n} \parallel \mathbf{z}$
the relation between ${\hat{ \mathbf{k}}}_{\textrm{in}}$  and
${\hat{ \mathbf{k}}}_{\textrm{out}}$ needed to evaluate 
$\gamma_B ({\hat{ \mathbf{k}}}_{\textrm{in}})$ 
in Eq. (\ref{CohFunc}) is given 
by $\phi_{\textrm {in}}=\phi_{\textrm {out}}$ and 
$\theta_{\textrm {in}}=\pi-\theta_{\textrm {out}}$, and the angle of incidence
is $\alpha=\theta_{\textrm {out}}$. 
In this situation, we find immediately that 
$\Delta_{\pm}({\hat{ \mathbf{k}}}_{\textrm{in}})=\Delta_{\pm}({\hat{ \mathbf{k}}}_{\textrm{out}})$
and $\sigma_{\pm}({\hat{ \mathbf{k}}}_{\textrm{in}})=\sigma_{\pm}({\hat{ \mathbf{k}}}_{\textrm{out}})$,
which directly yields
$\gamma_S ({\hat{ \mathbf{k}}}_{\textrm{out}})=\gamma_B ({\hat{ \mathbf{k}}}_{\textrm{out}})$.
Hence, both $\gamma_S$ and $\tilde{\gamma}_S$ are finally given by the bulk coherence functions 
for the \textit{same} $\hat{\mathbf{k}}$-direction, and therefore
the resulting angular resolved density of states $N(E,\hat{\mathbf{k}})$  
is identical to the one of the bulk.
No sub-gap Andreev bound states are generated by surface scattering. 
For small but finite energies the shape of  the local density of states 
is, however, determined by the actual nodal topology of  the gap functions
given in Eq. (\ref{Delta12}). This behavior is then
passed on to the resulting tunneling conductances, which
are presented in Fig. \ref{Fig01} for different values
of the ratio $ q= \psi / \Delta $.
Starting with $q=0$, we have a pure  $p$-wave state
with point nodes on the poles of the Fermi surface.
This leads to a parabolic behavior for low bias voltage reflecting the $E^2$-dependence
of the density of states. 
When $q$ is increased, the gap function $\Delta_+$ of Eq. (\ref{Delta12}) becomes
fully gapped (albeit anisotropically), while $\Delta_-$ exhibits immediately line nodes developing 
around the top and bottom of the Fermi sphere. 
Accordingly, a linear dependence can be observed for
low bias. For  $q=1$ the line node of $\Delta_-$ has moved to 
the equator of the Fermi sphere. This removes the linear low-bias behavior in $G$, since the nodal 
quasiparticles traveling parallel to the surface 
cannot contribute to the conductance Eq. (\ref{GFormel}) any longer.
As soon as $q> 1$ however, 
also  the gap function $\Delta_-$ is nodeless, which  results in 
the opening of a corresponding gap in the conductance
situated between 
the normalized energies $E/\Delta = \pm|q-1| $.  

\begin{figure}[t]
\includegraphics[width=0.85 \columnwidth]{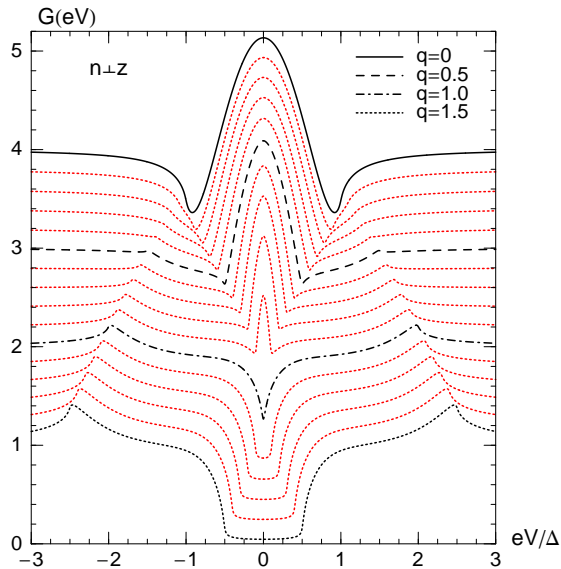}
\caption{\label{Fig03}
Normalized low temperature conductance $G(eV)$ for a 
normal metal/non-centrosymmetric superconductor junction
with boundary orientation $\mathbf{n} \perp \mathbf{z}$. 
The ratio $q$ of the gap amplitudes is varied from $0$ to $1.5$ in 
steps of $0.1$. Between two subsequent curves, a vertical  offset
of $0.2$ has been used for clarity of the plot.
The transparency of the interface is set to $D_0=0.1$. 
}
 \end{figure}

Next we turn to the boundary with normal vector $\mathbf{n} \perp \mathbf{z}$ 
where we can choose
freely $\mathbf{n} \parallel \mathbf{y}$. Here, the relation between
incoming and outgoing quasiparticle directions is
determined by $\phi_{\textrm {in}}=-\phi_{\textrm {out}}$ and $\theta_{\textrm {in}}=\theta_{\textrm {out}}$.
Furthermore, the angle of incidence is 
$\alpha=\arccos(\sin\phi_{\textrm {out}} \sin \theta_{\textrm {out}})$.
In contrast to $\mathbf{n} \parallel \mathbf{z}$, low-energy 
Andreev bound states can exist  for this orientation. By examination 
 of the angular resolved local density of states 
$N(E,\hat{\mathbf{k}})$ given in Eq. (\ref{NAngular}) we find the condition 
that
these bound states only occur for $\hat{\mathbf{k}}$-vectors 
with $\Delta_{-}<0$, i.e. the contributions  come from
those parts of the Fermi surface, where the $p$-wave
is stronger than the $s$-wave ($\sin \theta>q$). 
Actually, for every angle $\theta$ 
with $\Delta_- <0$ there is a whole branch
of bound states living between the energies $\pm |\Delta_{-}|$ 
determined by
\begin{equation}
\label{Branch}
\cos \phi =\pm \frac{E^2-\Delta_+ \Delta_{-}-\sqrt{\Delta_+ ^2-E^2} \sqrt{\Delta_{-} ^2-E^2}}{E(\Delta_+-\Delta_{-})}.
\end{equation} 
Zero-energy Andreev bound states occur for 
quasiparticle directions with $\phi=\pi/2$ which corresponds to vanishing
transverse momentum along the $x$-direction.
An example of the low-energy Andreev bound states in 
the plane $\theta=\pi/2$ for $q=0.5$ is provided by Fig. \ref{Fig02}.

The normalized low-temperature conductance for this situation is
presented in Fig. \ref{Fig03}. We find an extended zero-bias
conductance peak for $q=0$, analogous  to 
the one for a pure $p$-wave superconductor with
a cylindrical Fermi surface \cite{Yamashiro,Honerkamp}.  
This peaked structure at low bias voltages persists as long as  $q<1$
but becomes progressively narrower for $q \rightarrow 1$. These structures are easily understood
in the context of the Andreev bound states. 
For a given $q<1$ there is always a region around the equator of the Fermi sphere
with $\sin \theta >q$, where $\Delta_-<0$. Quasiparticles from those
parts of the Fermi surface support low-energy Andreev bound states
up to normalized energies of $\pm|\sin \theta-q|$ according to Eq. (\ref{Branch}).
The maximum possible energy $\pm|1-q|$ originates
from trajectories at the equator and corresponds to the width of the 
zero-bias anomalies
shown in Fig. \ref{Fig03}.  Moreover,
the total weight around zero bias naturally decreases when $q$ is increased, 
because the region on the Fermi sphere with $\Delta_- <0$ is reduced
and accordingly the contributions to Andreev bound states.
If $q\geq1$, however, $\Delta_{-}\geq0$ everywhere on the 
Fermi surface and Andreev bound states do not exist anymore. 
Therefore, the zero-bias anomaly disappears in the tunneling conductance. 

The results presented above have been achieved for a typical
tunnel junction with small interface transparency $D_0=0.1$.
Tunnel junctions with other interface transparencies 
chosen not too large to be consistent with the validity of Eq. (\ref{GFormel})
show the same characteristic behavior: 
For boundary orientations $\mathbf{n} \perp \mathbf{z}$ and 
$q<1$ the tunneling conductances exhibit a strong zero-bias anomaly.

Our discussion shows that the tunneling characteristics could give important information on
the gap structure of non-centrosymmetric superconductors, in particular in view of
the parity-mixing. Important is also the direction dependence which would provide
an important test for the theoretical picture developed for CePt$_3$Si sofar. It has also
been anticipated that the ratio $ q $ could be temperature dependent. Thus it is possible that
the tunneling spectrum displays qualitative changes as a function of temperature.

We would like to thank D.F. Agterberg, P.A. Frigeri, K. Wakabayashi and Y. Yanase for
helpful discussions. This work was financially supported by the Swiss Nationalfond and the
NCCR MaNEP.

\end{document}